\documentclass[aps,showpacs,superscriptaddress]{revtex4}
\usepackage{graphicx}
\begin {document}
\title{Nanomechanical-resonator-assisted
induced transparency in a Cooper-pair-box
system}
\author{Xiao-Zhong Yuan}
\affiliation{Department of Physics and Center for Theoretical
Sciences, National Taiwan University, Taipei 10617, Taiwan}
\affiliation{Department of Physics, Shanghai Jiao Tong
University, Shanghai 200240, China}
\author{Hsi-Sheng Goan}
\email{goan@phys.ntu.edu.tw}
\affiliation{Department of Physics and Center for Theoretical
Sciences, National Taiwan University, Taipei 10617, Taiwan}
\affiliation{Center for Quantum Science and Engineering, National Taiwan University, Taipei 10617, Taiwan} 
\author{Chien-Hung Lin}
\affiliation{Department of Physics and Center for Theoretical
Sciences, National Taiwan University, Taipei 10617, Taiwan}
\author{Ka-Di Zhu}
\author{Yi-Wen Jiang}
\affiliation{Department of Physics, Shanghai Jiao Tong
University, Shanghai 200240, China}

\begin{abstract} We propose a scheme to demonstrate the
electromagnetically induced transparency (EIT) in a system of a
superconducting Cooper-pair box coupled to a nanomechanical
resonator. In this scheme, the nanomechanical resonator plays an important
role to contribute additional auxiliary energy levels to the
Cooper-pair box so that the EIT phenomenon could be realized in
such a system. 
We call it here resonator-assisted induced transparency (RAIT). 
This RAIT technique provides a detection scheme in a
real experiment to measure physical properties, such as the
vibration frequency and the decay rate, of the coupled
nanomechanical resonator.
\end{abstract}
\pacs{85.85.+j, 85.25.-j, 42.50.Gy}

\keywords{Nanomechanical-resonator, Electromagnetically induced
transparency, Cooper-pair box}
\maketitle


\section{Introduction}
To be able to observe quantum phenomena in mesoscopic
(macroscopic) physical systems which contain many millions of
atoms is of great importance in quantum mechanics and also quantum
information science. Currently, there are experimental and
theoretical groups devoting to observing quantum effects in truly
solid-state mechanical oscillator or cantilever systems
\cite{Bocko96,Roukes,Blencowe04,Schwab05, Cleland, Schwab,
Schwab06, Marshall, Mancini, Goan04, Goan05, Bose06, Peano06,
Pirandola, Xue07,Plenio02, Plenio04, Eisert, Goan07, Jacobs07,
Clerk07,Buks08}. In the regime when the individual mechanical quanta are
of the order or greater than the thermal energy, quantum effects
become important and the motion of the mechanical resonator or
cantilever is close to or on the verge of the quantum limit.
Coupling such a mechanical system to an electrical motional
transducer might enable us to observe and control the motional
quantum state of the mechanical system.

A Superconducting Cooper-pair box (CPB)
\cite{Nakamura99,Makhlin01,Astafiev04,Duty04,You05,Clarke08}
consists of a superconducting island (box) weakly linking to a
superconducting reservoir by a Josephson tunnel junction. The
coherent controls and manipulations of the effective two-level
quantum system of a CPB (a charge qubit) have been demonstrated
\cite{Nakamura99,Astafiev04,Duty04}. Due to its controllability,
a CPB has been proposed to be one of ideal candidates to act as a
transducer or a mediator to couple to a nanomechanical resonator
(NR). With appropriate quantum state controls of a CPB, the CPB
has been proposed to be used to produce desired NR Fock state and
perform a quantum non-demolition (QND) measurement of  NR Fock
state \cite{Irish03},  to drive a NR into a superposition of
spatially separated states and probe their decoherence rate
\cite{Armour}, to cool the NR to its ground state and generate
squeezed state of the NR \cite{Martin04}, to probe the quantum
mechanical feature of tiny motions of a NR \cite{Wei06}, to
integrate a superconducting transmission line resonator with a
NR \cite{Sun06}, and to demonstrate progressive quantum decoherence
\cite{Sun04}.

Recently, a high-frequency mechanical resonator beam that operates
in the GHz range has been reported \cite{Huang03}. For a NR
operating at the fundamental frequency of GHz and at a temperature
of 10-100mK, some interesting phenomena close to or on the verge
of the quantum limit may be observed. In this paper, we propose a
scheme to demonstrate the electromagnetically induced transparency
(EIT) in a system of a superconducting CPB qubit coupled to a NR.
The conventional EIT effect occurs in an ensemble of three-level
$\Lambda$-type atomic system with two lower states coupled
respectively to an excited state with two laser fields (control
and probe fields) \cite{Harris97, Fleischhauer05}. A typical
experiment is conducted by scanning the probe laser frequency and
measuring its transmitted intensity. The transparency of the
medium takes place when the absorption on both transitions is
suppressed due to destructive interference between excitation
pathways to the common upper level. In addition to the absorption
eliminated via quantum interference, the EIT effect has also been
shown to be an active mechanism to slow down or stop light pulse
completely in various systems, such as ultracold gas of sodium
atoms\cite{Harris99}, rare-earth-ion-doped crystals \cite{Ham97},
semiconductor quantum wells \cite{Serapiglia00}, quantum dot
exciton systems \cite{Zhu06}, and systems with four-lever or
multi-level cells \cite{Knight02}. Recently, it has been proposed
to use a superconductive analogy to EIT in a persistent-current flux
qubit biased in a $\Lambda$ configuration to probe small qubit
errors due to decoherence or imperfect state preparation
\cite{Orlando04, Orlando06}. Here we show that the EIT phenomenon
could be realized in an effective two-level superconducting CPB
charge qubit coupled to a NR. 
The capacitive CPB-NR coupling plays an
important role to contribute additional auxiliary energy levels 
for EIT to occur.
As a result, this resonator-assisted induced transparency (RAIT) technique
provides a detection scheme in a real experiment to measure
physical properties, such as the vibration frequency and the decay
rate of the coupled NR, or the decay and decoherence
rates of the CPB qubit, if one set of the values of either the CPB
or the NR properties is known by other means.

\section{Model and Calculations}

\begin{figure}[htbp]
\begin{center}
\includegraphics[width=8.5cm] {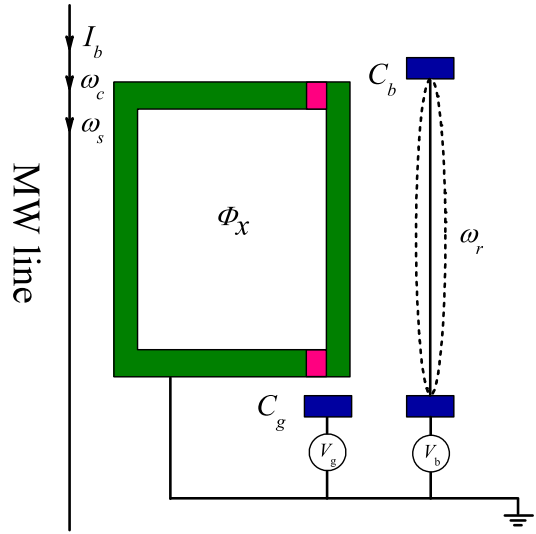}
\end{center}
\caption{Schematic diagram of a nanomechanical resonator (NR)
coupled to a Cooper-pair box (CPB) qubit. 
The microwave currents (with frequencies $\omega_s$, $\omega_c$)
and a direct current ($I_b$) are applied to flow along the MW line 
beside the CPB
to control the magnetic flux through the CPB loop.}
\label{fig:setup}
\end{figure}

In our model, we assume that a nanomechanical-resonator (NR)
couples capacitively to a Cooper-pair box (CPB) qubit 
(see Fig.~\ref{fig:setup}). The
tunnel junction of the CPB shown in Fig.~\ref{fig:setup} is split
into two to form a SQUID loop  \cite{Nakamura99,Astafiev04} which
allows us to control its effective Josephson energy with a small
external magnetic field or magnetic flux. 
Two microwave currents are applied in a microwave (MW) line \cite{Mooij03} 
beside the CPB to induce oscillating magnetic fields
in the Josephson junction SQUID loop of the CPB qubit. We call one of them the
control current with frequency $\omega_c$ and amplitude ${\cal
E}_c$. The other is called the signal (probe) current with
frequency $\omega_s$ and amplitude ${\cal E}_s$. In addition, a
direct current $I_b$ is also applied to the MW line to control the magnetic flux
through the SQUID loop and thus the effective Josephson coupling
of the CPB qubit. The Hamiltonian of the total system can be
written as \cite{Irish03, Sun06}:
\begin{eqnarray}
H&=&H_{CPB}+H_{NR}+H_{int},\\
H_{CPB}&=&\hbar\omega_qS_z-\hbar E_J\cos\left[\frac{\pi\phi_x(t)}{\phi_0}\right]S_x,\label{HCPB}\\
H_{NR}&=&\hbar\omega_r a^+a,\\
H_{int}&=&2\hbar\lambda(a^++a)S_z.
\end{eqnarray}
Here $H_{CPB}$ and $H_{NR}$ are respectively the Hamiltonians of
the CPB qubit and the NR. $H_{int}$ is the interaction between
them \cite{Irish03}. The operators $a$, $a^+$ denote the creation
and annihilation operators for the NR with frequency $\omega_r$
and mass $m$. The two-level system of the CPB qubit can be
characterized by the pseudospin-1/2 operators $S_z$ and
$S_x=S_++S_-$. $\omega_q=4E_C(2n_g-1)$ is the electrostatic energy
and $E_J$ is the maximum Josephson energy. Here,
$E_C=e^2/(2C_\Sigma)$ is the charging energy with $C_\Sigma$ the
total CPB capacitance and $n_g=(C_bV_b+C_gV_g)/(2e)$ is the
dimensionless gate charge, where $C_g$ and $V_g$ are, respectively, the gate
capacitance and gate voltage of the CPB qubit, and $C_b$ and $V_b$
are, respectively, the capacitance and voltage between the NR and
the CPB island. The capacitive interaction strength can be written as
$\lambda=4E_CC_gV_g/(2de\sqrt{2\hbar m\omega_r})$, where $d$ is the
distance between the NR and the CPB  \cite{Irish03}. The coupling
between the MW line and the CPB qubit in the second term of
Eq.~(\ref{HCPB}) results from the total externally applied
magnetic flux $\phi_x(t)=\phi_q(t)+\phi_b$ through the CPB qubit
loop of an effective area $S$ with $\phi_0=\hbar/(2e)$ being the flux
quantum \cite{Sun06}. Here
\begin{equation}
\phi_q(t)=\mu_0S\,I(t)/(2\pi r),
\end{equation}
 with $r$ being the
distance between the MW line and the qubit, and $\mu_0$ being the vacuum
permeability \cite{Sun06}. $\phi_q(t)$ and $\phi_b$ are
produced, respectively, by the microwave current
\begin{equation}
  \label{eq:current}
I(t)={\cal E}_c\cos(\omega_ct)+{\cal E}_s\cos(\omega_st+\delta')
\end{equation}
 and the direct current $I_b$ in the MW line.  For simplicity, we
assume the phase factor $\delta'=0$ as it is not difficult to show that the results of this paper do not depend on the value of $\delta'$.

Choosing the direct current $I_b$ and the microwave current $I(t)$
such that $\phi_b\gg\phi_q(t)$ and $\pi\phi_b/\phi_0=\pi/2$, 
we have
\begin{eqnarray}
\hbar E_J\cos\left[\frac{\pi\phi_x(t)}{\phi_0}\right]\approx-\hbar
E_J\frac{\pi\phi_q(t)}{\phi_0}. \label{fluxapprox}
\end{eqnarray}
We can work in a frame rotating at the frequency
$\omega_c$ of the control current
 and the total Hamiltonian in this rotating frame becomes
\begin{eqnarray}
H_c&=&\hbar\Delta S_z+\hbar\omega_r
a^+a+2\hbar\lambda(a^++a)S_z\nonumber\\
&&+\hbar\Omega(S_++S_-)+\mu {\cal E}_s(S_+e^{-\textmd{i}\delta
t}+S_-e^{\textmd{i}\delta t}).
\label{HrotWc}
\end{eqnarray}
In analogy to the case of a two-level atom driven by bichromatic
electromagnetic waves, here
\begin{equation}
\mu=\mu_0S\hbar E_J/(4r\phi_0)
 \label{eq:dipole_moment}
\end{equation}
is the effective  ``electric dipole moment'' of the qubit,
\begin{equation}
  \label{eq:Rabi}
  \Omega=\mu{\cal E}_c/\hbar
\end{equation}
is the effective ``Rabi frequency'' through the control current,
\begin{equation}
\label{eq:Delta}
\Delta =\omega_{q}-\omega_{c} 
\end{equation}
is the detunning between the CPB qubit resonance frequency and control current frequency, and
\begin{equation}
  \label{eq:detuning_current}
  \delta=\omega_s-\omega_c
\end{equation}
is the detuning between the signal (probe) current frequency and the
control current frequency.

Furthermore, we may take into account the decoherence and
relaxation of the CPB qubit and NR by including their coupling
to external environments into the Hamiltonian \cite{Gardiner00, Milburn94,
Carmichael99, Breuer02}. We assume that the environments to which
the CPB and NR couple respectively could be described as
independent ensembles of harmonic oscillators with respective
spectral densities characterizing their properties. We also assume
that the NR interacts bilinearly with the environment through
their position operators, and the CPB interacts through $S_x$
operator and $S_z$ operator with the environment position
operators. The $S_x$ coupling to the environment models the
relaxation (and thus also decoherence) process of the CPB qubit,
while the $S_z$ coupling to the environment models the pure
dephasing process of the CPB qubit \cite{Gardiner00, Milburn94,
Carmichael99, Breuer02}. Generally speaking, the value of $\omega
_{q}$ of the CPB qubit is considerably greater than the value of
$\omega_r$ of the NR. Therefore, it may be plausible to apply the
rotating-wave approximation to the CPB-environment coupling term,
but not for NR-environment coupling term in the system-environment
interaction Hamiltonian. By following the standard procedure
\cite{Gardiner00, Milburn94, Carmichael99, Breuer02}, it is then
straightforward to derive the Born-Markovian master equation of
the reduced density matrix of the CPB-NR system, $\rho(t)$,
through tracing out the environmental degrees of freedom as:
\begin{eqnarray}
\frac{d{\rho }}{dt} &=&-\frac{i}{\hbar }\left[ H_c,\rho \right] +A\left\{ %
\left[ S_{-},\left[ S_{+},\rho \right] \right] +h.c.\right\} +B\left\{
[S_{-},\{S_{+},\rho \}]+h.c\right\}   \nonumber \\
&&+E\left[ S_{z},\left[ S_{z},\rho \right] \right] +D\left[ Q,\left[ Q,\rho %
\right] \right] +G\left[ Q,\left[ P,\rho \right] \right] +\frac{i}{\hbar }L%
\left[ Q,\left\{ P,\rho \right\} \right] ,  \label{mastereq1}
\end{eqnarray}%
where $Q=-2\lambda (a^{+}+a)$ and $P$ are the position and momentum
operators of the NR, respectively. The coefficients $A,B,E,D,G$ and $L$ are
related to the characteristics of the coupling, and to the structure and
properties of the environments. Their explicit forms can be written as
\begin{eqnarray}
A &=&-\frac{1}{2\hbar }\left\{ \frac{\gamma _{1}}{2}\left( 1+2N\left( \omega
_{q}\right) \right) \right\} , \\
B &=&\frac{1}{2\hbar }\left\{ \frac{\gamma _{1}}{2}\right\} ,
\\
E &=&-\frac{1}{\hbar }\left\{ \frac{\gamma _{2}}{2}\left( 1+2N\left(
0\right) \right) \right\} ,   \\
D &=&-\frac{1}{4\hbar }\gamma _{3}\left( 1+2N\left( \omega _{r}\right)
\right) ,  \\
G &=&-\frac{1}{2\hbar }\frac{1}{m\omega _{r}}\Delta _{3},   \\
L &=&-\frac{1}{4m\omega _{r}}\gamma _{3},
\end{eqnarray}%
where%
\begin{eqnarray}
\gamma _{1} &=&2\pi J_{x}\left( \omega _{q}\right) , \\
\gamma _{2} &=&2\pi J_{z}\left( 0\right) ,   \\
\gamma _{3} &=&2\pi J_{R}\left( \omega _{r}\right) ,  \\
\Delta _{3} &=&\mathcal{P}\int_{0}^{\infty }d\omega \frac{J_{R}\left( \omega
\right) }{\omega -\omega _{r}}\left( 1+2N\left( \omega \right) \right) .
\end{eqnarray}%
Here, $J_{x}$, $J_{z}$ and $J_{R}$ are the spectral densities of the
respective environments coupled through $S_{x}$ and $S_{z}$ to the CPB, and
through $Q$ to the NR, respectively. $N\left( \omega \right) =1/[\exp (\hbar
\omega /{k_{B}T})-1]$ is the Boltzman-Einstein distribution of the thermal
equilibrium environments. $\mathcal{P}$ denotes the principal value of the
argument. Note that $\omega_r$ and $\omega_q$ in $H_c$ in Eq.~(\ref{mastereq1}) should be regarded as the real physical frequencies in which the renormalization and Lamb shifts due to the interactions with the
environments have been included.
With the master equation (\ref{mastereq1}), we can obtain the
equation of motion for the mean (or expectation value) of any physical
operation $O$ of the CPB-NR system by calculating $\langle \dot{O}\left(
t\right) \rangle =Tr\left[ O\dot{\rho}\left( t\right) \right] $. For
convenience, in the following we denote variable $O(t)$ as its expectation
value $\langle O(t)\rangle $, and will clarify its definition when there is
a potential confusion. We thus have
\begin{eqnarray}
&&\frac{dS_{-}}{dt}=\left[ -\frac{1}{T_{2}}-i(\Delta +Q)\right]
S_{-}+2i\Omega S_{z}+2\frac{i}{\hbar }\mu S_{z}\mathcal{E}_{s}e^{-i\delta t},
\label{eq:S-} \\
&&\frac{dS_{z}}{dt}=-\frac{1}{T_{1}}\left( S_{z}+\frac{1}{2}\right) -i\Omega
(S_{+}-S_{-})-i\frac{\mu }{\hbar }(S_{+}\mathcal{E}_{s}e^{-i\delta t}-S_{-}%
\mathcal{E}_{s}^{\ast }e^{i\delta t}), \\
&&\frac{d^{2}Q}{dt^{2}}+\gamma \frac{dQ}{dt}+\omega _{r}^{2}Q=-8\omega
_{r}^{3}\lambda _{0}S_{z},  \label{eq:Q}
\end{eqnarray}%
where
\begin{equation}
\lambda _{0} =\lambda ^{2}/\omega _{r}^{2}.
\label{lambda0}
\end{equation}
Please note that the variables $S_-$, $S_z$ and $Q$ in
Eqs.~(\ref{eq:S-})-(\ref{eq:Q}) stand for the expectation values
of their corresponding operators, respectively, except that the
product of the variables $Q$ and $S_-$ in Eq.~(\ref{eq:S-}) should be
regarded as $\langle Q\, S_-\rangle$. The decoherence time $T_{2}$ 
and excited-state relaxation time $T_{1}$ of the CPB qubit,
and the decay rate $\gamma $
of the NR are derived microscopically as%
\begin{eqnarray}
T_{2}& =&\left[ \frac{1}{\hbar}\left\{ \frac{\gamma _{1}}{2}\left(
1+2N\left( \omega _{q}\right) \right) \right\} +\frac{4}{\hbar
}\left\{ \frac{\gamma _{2}}{2}\left( 1+2N\left( 0\right) \right)
\right\} \right]
^{-1}, \label{T2}\\
T_{1}& =&\left[ \frac{2}{\hbar}\left\{ \frac{\gamma _{1}}{2}\left(
1+2N\left( \omega _{q}\right) \right) \right\} \right] ^{-1}, 
\label{T1}\\
\gamma & =&\frac{1}{2m\omega _{r}}\gamma _{3}.
\end{eqnarray}%
Note that if the pure dephasing coupling were dropped, i.e., $\gamma_2=0$, then $T_2=2T_1$. This is often the case for a two-level atomic system.
After defining $p=\mu S_{-}$, $k=2S_{z}$, we have
\begin{eqnarray}
&&\frac{dp}{dt}=\left[ -\frac{1}{T_{2}}-i(\Delta +Q)\right]
p+i\frac{\mu
^{2}k {\cal E}}{\hbar }, \label{pEOM}\\
&&\frac{dk}{dt}=-\frac{1}{T_{1}}\left( k+1\right) -4\frac{\text{Im}%
(p{\cal E}^{\ast })}{\hbar }, \label{kEOM} \\
&&\frac{d^{2}Q}{dt^{2}}+\gamma \frac{dQ}{dt}+\omega_r
^{2}Q=-4\lambda _{0}\omega_r^{3}k, \label{QEOM}
\end{eqnarray}%
where ${\cal E}={\cal E}_c+{\cal E}_se^{-i\delta t}$ and ${\cal
E}_c=\hbar\Omega/\mu$. Note that the $Q\, p$ term in
Eq.~(\ref{pEOM}) should be regarded as the expectation value of
$\langle Q\,p\rangle$. The above equations can not be solved
as they are not closed.

In order to solve these equations, we first take the semiclassical
approach by factorizing the NR and CPB qubit degrees of freedom,
i.e., $\langle Q\,p\rangle=\langle Q\rangle\langle p\rangle$. This
ignores any entanglement between these systems.
We will study the EIT behavior in the context of a weak signal
(probe) current amplitude ${\cal E}_s$ in the presence of a strong
control current amplitude ${\cal E}_c$. To obtain analytical
solution, we make the ansatz
\begin{eqnarray}
p &=&p_{0}+p_{1}e^{-i\delta t}+p_{-1}e^{i\delta t}, \label{ansatz_p}\\
k &=&k_{0}+k_{1}e^{-i\delta t}+k_{-1}e^{i\delta t}, \\
Q &=&Q_{0}+Q_{1}e^{-i\delta t}+Q_{-1}e^{i\delta t}.
\label{ansatz_Q}
\end{eqnarray}%
By substituting Eqs.~(\ref{ansatz_p})-(\ref{ansatz_Q}) into
Eqs.~(\ref{pEOM})-(\ref{QEOM}) and working to the lowest order in
${\cal E}_s$, we finally obtain in the steady-state the following
solutions
\begin{eqnarray}
k_{1} &=&\frac{2\frac{T_{1}}{T_{2}}\Omega_{c}^{2}k_{0}\theta }{\frac{T_{1}}{
T_{2}} i \delta _{0}-1-2\frac{T_{1}}{T_{2}}\beta }\cdot \frac{{\cal E}_s}{{\cal E}_c}, \\
p_{1} &=&\frac{\frac{i4\lambda _{0}k_{1}\omega_r^{3}}{-\delta
^{2}-i\gamma
\delta +\omega_r^{2}}\cdot \frac{\mu ^{2}k_{0}{\cal E}_c}{\Delta -4\lambda_0\omega_r k_0-\frac{i}{%
T_{2}}}+i\mu^{2}(k_{0}{\cal E}_s+k_{1}{\cal E}_c)}{i\hbar (\Delta -4\lambda_0\omega_r k_0-\delta )+%
\frac{\hbar }{T_{2}}},
\end{eqnarray}%
where
\begin{eqnarray}
\theta  &=&\frac{1}{i(\Delta _{c}-4\lambda _{0}\omega
_{0}k_{0}-\delta
_{0})+1}+\frac{i}{\Delta _{c}-4\lambda _{0}\omega _{0}k_{0}+i}, \\
\beta  &=&\frac{4\lambda _{0}\omega _{r}\eta \frac{\Omega _{c}^{2}k_{0}}{%
\Delta _{c}-4\lambda _{0}\omega _{0}k_{0}-i}+\Omega
_{c}^{2}}{i(\Delta _{c}-4\lambda _{0}\omega _{0}k_{0}-\delta
_{0})+1}+\frac{4\lambda _{0}\omega _{0}\eta \frac{\Omega
_{c}^{2}k_{0}}{\Delta _{c}-4\lambda _{0}\omega _{0}k_{0}+i}+\Omega
_{c}^{2}}{-i(\Delta _{c}-4\lambda _{0}\omega _{0}k_{0}+\delta
_{0})+1}, \\
\eta &=&\frac{\omega _{0}^2}{\omega
_{0}^2-i\gamma_0\delta_0-\delta_0^2},\
\end{eqnarray}%
and dimensionless variables $\omega_{0}=\omega_r T_{2}$,
$\gamma_{0}=\gamma T_{2}$, $\delta_{0}=\delta T_{2}$, $\Omega
_{c}=\Omega T_{2}$, and $\Delta _{c}=\Delta T_{2}$ are introduced
for later numerical convenience. The zero-order population
inversion of the CPB is determined by the following equation:
\begin{eqnarray}
(k_0+1)[(\Delta_c-4\lambda_0\omega_0k_0)^2+1]+4\frac{T_1}{T_2}\Omega_c^2k_0=0.
\label{eq:zero-order}
\end{eqnarray}
The cubic Eq.~(\ref{eq:zero-order}) has either a single or three
real roots. The latter case just corresponds to the intrinsic
bistable states which we will not discuss here.

To observe the EIT phenomena, we calculate the absorption power of
the signal (probe) current as a function of its frequency. We note
that the second term containing $S_x$ in Eq.~(\ref{HCPB})
describing the Cooper-pair tunneling energy controlled by the
external flux which is induced by the applied MW line current. By
using the approximation of Eq.~(\ref{fluxapprox}), the absorption
power of the signal current can be expressed as
\begin{eqnarray}
P_{abs}=\pi\frac{\hbar E_J}{\phi_0}S_x\frac{d\phi_s}{dt},
\end{eqnarray}
where $\phi_s=\mu_0SI_s/(2\pi r)$ is the magnetic flux produced by
the signal current $I_s={\cal E}_s\cos(\omega_st)$.
There are several terms in $S_x$ as indicated in
Eq.~(\ref{ansatz_p}) which is written in the frame rotating at the
control current frequency $\omega_c$. It is obvious that only the
terms $p_1e^{-i\omega_st}$ and $p_1^*e^{i\omega_st}$  in the
laboratory frame have non-zero contributions to the time-averaged
absorption power which can be measured in experiments. In this
way, after the time average, we have
\begin{eqnarray}
P_{abs}=-2\mu {\cal E}_s\omega_s\text{Im}(p_1), \label{abs_power}
\end{eqnarray}
where Im denotes taking the imaginary part, and $\mu=\mu_0S\hbar
E_J/(4r\phi_0)$.

\section{Experimental parameters and results}
We discuss how the EIT phenomena could be designed and realized in
the CPB-NR system with realistically reasonable parameters. The
typical values of the charging energy $E_C$ and the Josephson
coupling energy $E_J$ of a CPB charge qubit are often designed such that
$E_C\gg E_J$, so $E_C=40$ $\textmd{GHz}$ and $E_J=2$ $\textmd{GHz}$
are chosen. The values $\lambda=50$ $\textmd{MHz}$ and $\omega_r=1$ 
$\textmd{GHz}$ are used as in Refs. \cite{Irish03, Armour, Wei06,
Schwab05}. The decay (relaxation) rate and decoherence rate of the
CPB system depend on temperatures and the qubit operational points 
which could be
controlled by the external gate voltage and magnetic flux. It has been
reported that the relaxation rate $1/T_1=4$ $\textmd{MHz}$
\cite{Martin04, Lehnert03} and the decoherence time $T_2$ can
reach the order of microsecond at the degeneracy point
\cite{Vion02}. In our system, the CPB qubit may not be tuned at the
degeneracy point, so it could be sensitive to the
inevitable charge noise that is present in the circuits. Hence we
choose the decoherence rate conservatively to be
$1/T_2=20$ $\textmd{MHz}$. The dominant mechanism for the damping
of the resonator mode may come from the coupling to the phonon
modes of the support, which could lead to the decay rate
$\gamma=0.01$ $\textmd{MHz}$ \cite{Martin04}. For numerical
convenience, we use dimensionless quantities for these quantities
as follows. $\gamma_0=\gamma T_2=5\times10^{-4}$,
$\omega_0=\omega_r T_2=50$,
$\lambda_0=\lambda^2/\omega_r^2=2.5\times10^{-3}$. For $S=1$ $\mu
\textmd{m}^2$, $r=1$ $\mu \textmd{m}$, and ${\cal E}_c=200$ $\mu
\textmd{A}$, we have $\mu/\hbar=\mu_0S E_J/(4r\phi_0)\approx300$
$\textmd{GHz/A}$, $\Omega_c=\Omega T_2=(\mu/\hbar){\cal
E}_cT_2=3$.

\begin{figure}[htbp]
\begin{center}
\includegraphics  [height=5cm,width=9cm] {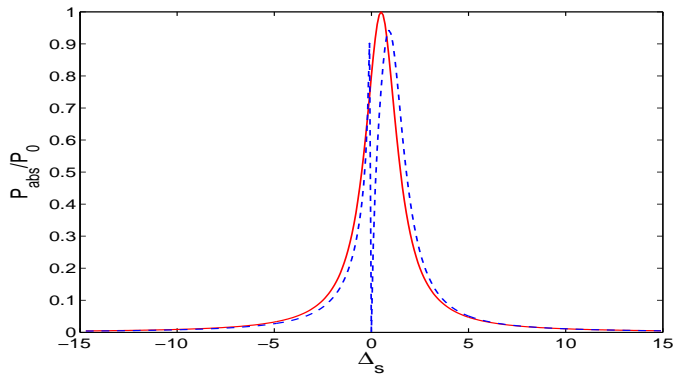}
\end{center}
\caption{Scaled absorption power profiles of the signal current as a
function of the detuning $\Delta_s$ for $\Omega_c=0$ (solid curve) and $\Omega_c=3$ (dashed curve). Other parameters used are
 $\lambda_0=2.5\times10^{-3}$, $\Delta_c=\omega_0=50$, $\gamma_0=5\times10^{-4}$.}
\label{fig:absorption_Omega}
\end{figure}

\begin{figure}[htbp]
\begin{center}
\includegraphics  [height=5cm,width=9cm] {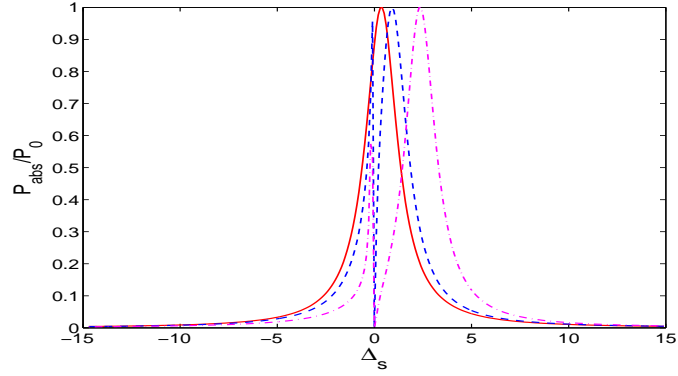}
\end{center}
\caption{Shifts of the absorption resonance profiles
for different values of the coupling constant $\lambda_0$
between the CPB and NR: $\lambda_0=0$ (solid
curve), $\lambda_0=2.5\times10^{-3}$ (dashed curve), and
$\lambda_0=0.01$ (dot-dashed curve).
Other parameters used are $\Omega_c=3$,
$\Delta_c=\omega_0=50$, $\gamma_0=5\times10^{-4}$.}
\label{fig:absorption_lambda}
\end{figure}

Figure \ref{fig:absorption_Omega} plots the absorption power of the signal current $P_{abs}$ of Eq.~(\ref{abs_power})
as a function of the detuning $\Delta_s$
($\Delta_s=(\omega_s-\omega_q)T_2$). The absorption power is
scaled in unit of $P_0$ which is the maximum value of $P_{abs}$
when $\Omega_c=0$. In the absent of the control current
($\Omega_c=0$), the solid curve shows a standard resonance
absorption profile of the signal current in the CPB system,
with the centre of the curve shifted from the resonance $\omega_s=\omega_q$
a bit. This is due to the coupling $\lambda_0$ between the CPB and NR
\cite{Irish03,Wei06}.
Furthermore, the resonant frequency shift
increases with the increase of the coupling constant $\lambda_0$.
When the control current is turned on ($\Omega_c=3$), a narrow
non-absorption hole appears at $\Delta_s=0$ as shown in the dashed
curve in Fig.~\ref{fig:absorption_Omega}. This indicates that the
signal current has a narrow peak of induced transparency (no
absorption).
Without the coupling $\lambda_0$ between the CPB and NR, such a
phenomenon will not appear. This can be seen from Fig.~\ref{fig:absorption_lambda} in which a standard absorption resonance profile for $\lambda_0=0$ (in solid line) appears. The central resonance peak shifts also a little bit from $\Delta_s=0$ (i.e., $\omega_s=\omega_q$) due to the presence of the control current ($\Omega_c=3$). When $\lambda_0\neq 0$ (dashed and  dot-dashed curves), the coupling prevents absorption of the signal (probe) current in a narrow portion of the resonance profile.
Therefore, we call it
resonator-assisted induced transparency (RAIT).  
As we have chosen $\Delta_c=\omega_0$, the minima
of the non-absorption holes (valleys) appear at $\Delta_s=0$ in
Fig.~\ref{fig:absorption_Omega} and Fig.~\ref{fig:absorption_lambda}.
Otherwise, they will move away from the point $\Delta_s=0$.
Recently, Radeonychev \textit{et al.} \cite{Radeonychev06} 
have shown that resonant
transparency could occur in a two-level quantum system induced via
mechanical (acoustical) harmonic vibration of a solid medium along the propagation of multi-frequency laser radiation. They concluded that the
atomic acoustical vibration plays a key role in this so-called  
acoustically induced transparency (AIT).

\begin{figure}[htbp]
\begin{center}
\includegraphics  [width=12cm] {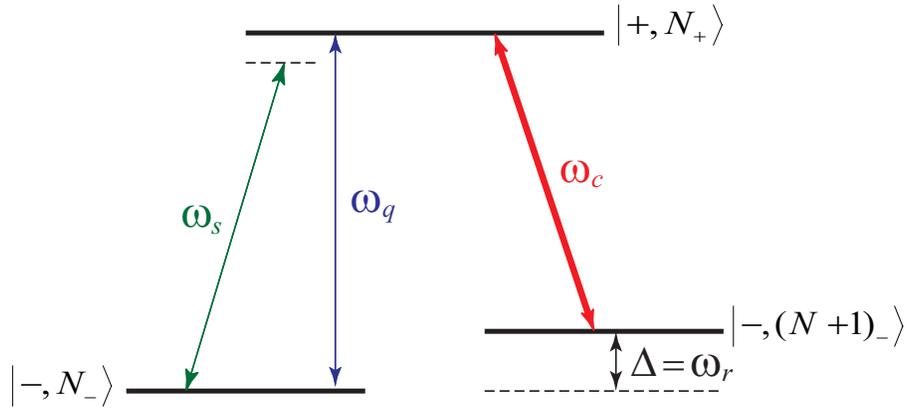}
\end{center}
\caption{Schematic illustration of quantum three-level structure in the coupled CPB-NR system. Here a strong control current of frequency $\omega_c$ is tuned on resonance between the $|+,N_+\rangle$ and $|-,(N+1)_-\rangle$ transition with $\Delta=\omega_q-\omega_c=\omega_r$, where $\omega_r$ is the resonant frequency of the NR and $\omega_q$ is the resonant frequency of the CPB qubit. A weak signal (probe) current with a tunable frequency $\omega_s$ is applied, and its absorption power profile near the resonant frequency of the CPB qubit is measured.}  
\label{fig:energy}
\end{figure}

Alternatively, we may also understand how the RAIT could
occur in the CPB and NR system as follows. 
We first consider the NR and CPB system without the application of the control and signal MW currents. The Hamiltonian, excluding the coupling to the external environments, from Eq.~(\ref{HrotWc}) is then
\begin{equation}
H_c=\hbar\omega_qS_z+\hbar\omega_r a^+a+2\hbar\lambda(a^++a)S_z.
\label{HCPBNR}
\end{equation}
The capacitive coupling between the CPB and the NR indicates that one charge state of the CPB will attract the NR and shift its equilibrium position near the CPB, while the other charge state will repel the CPB. Therefore, a displaced oscillator basis will allow us to use the ordinary harmonic oscillator formalism within each displaced potential well. The Hamiltonian of Eq.~(\ref{HCPBNR}) can then be diagonalized \cite{Irish05,Irish07} in the eigenbasis of 
\begin{equation}
  \label{eq:displaced_basis}
  |\pm,N_\pm\rangle= |\pm\rangle_z\otimes e^{\mp(\lambda/\omega_r)(a^+-a)}|N\rangle,
\end{equation}
with the eigenenrgies
\begin{equation}
  \label{eq:eigenenergy}
  E_\pm=\pm \hbar\omega_q+\hbar \omega_r (N-\lambda_0),
\end{equation}
where the CPB qubit states $|\pm\rangle_z$ are eigenstates of $S_z$ with  the excited state $|+\rangle_z=|e\rangle$ and the ground state $|-\rangle_z=|g\rangle$, the oscillator states $|N_\pm\rangle$ are position-displaced Fock states, and $\lambda_0$ is defined in Eq.~(\ref{lambda0}). 
Note that $|+,N_+\rangle$ and $|-,N_-\rangle$ form, respectively, an orthonormal basis with $\langle M_+|N_+\rangle=\delta_{MN}$ and $\langle M_-|N_-\rangle=\delta_{MN}$, but the states $|N_+\rangle$ and $|N_-\rangle$ are not mutually orthogonal and their inner products are given by \cite{Irish05,Irish07} 
\begin{equation}
  \label{eq:inner_product}
  \langle M_-|N_+\rangle=e^{-2\lambda_0}\left(\frac{2\lambda}{\omega_r}\right)^{N-M}\sqrt{\frac{M!}{N!}}\, L^{N-M}_{M}(4\lambda_0),
\end{equation}
where $L_i^j(y)$ is the associated Laguerre polynomial.
Thus the coupling to NR could provide the CPB qubit with additional auxiliary energy levels to realize the EIT phenomena. For the parameters used in the simulations, the quantum three levels that may realize the EIT phenomena could be chosen as $|-,N_-\rangle$, $|-,(N+1)_-\rangle$ and $|+,N_+\rangle$ illustrated in Fig.~\ref{fig:energy}.
In Fig.~\ref{fig:energy}, the state $|-,N_-\rangle$ has the same parity as state $|-,(N+1)_-\rangle$, and thus the effective electric-dipole transition is forbidden.  On the other hand, the state $|+,N_+\rangle$ is of opposite parity and thus has a non-zero effective electric-dipole coupling to both $|-,N_-\rangle$ and  $|-,(N+1)_-\rangle$ states. 
These conditions satisfy the specific restrictions on the configuration of the three levels (states) in atoms to realize EIT. That is, two of the three possible transitions between the states must be dipole-allowed, i.e. the transitions can be induced by an oscillating electric field. The third transition should be dipole-forbidden.  In our setup, the strong control field (current) is tuned on resonance between the $|+,N_+\rangle$ and $|-,(N+1)_-\rangle$ transition. The weak probe or signal current is tuned near resonance between the two states, $|+,N_+\rangle$ and $|-,N_-\rangle$. Then the singnal-current absorption power profile of the transition is measured. As in a conventional three-level $\Lambda$-type atomic system  \cite{Harris97, Fleischhauer05}, the strong control field current with the detuning equal to the vibration frequency of the NR drives the coupled CPB-NR system. As a result, the dressed CPB-NR coupled system becomes transparent for a weak signal current with a
frequency matching the resonant frequency of CPB qubit.
We note that the selection rule here is somewhat different from that of the superconducting flux-qubit circuit which shows a EIT phenomenon in a $\Lambda$ configuration \cite{Orlando04, Orlando06}. There, dipole-like coupling is allowed between all pairs of levels due to the symmetry breaking of the potential of the flux qubit \cite{Liu05}. 

\begin{figure}[htbp]
\begin{center}
\includegraphics  [height=5cm,width=9cm]{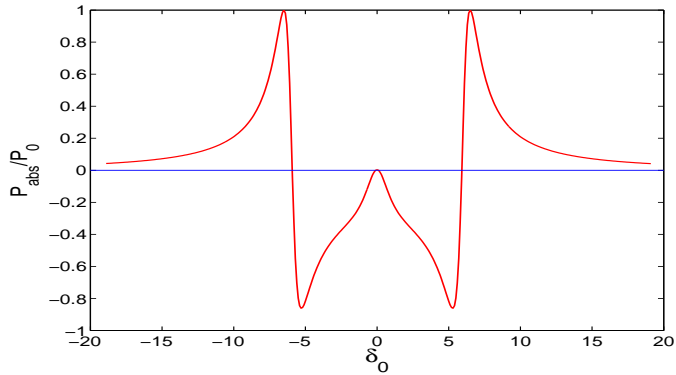}
\end{center}
\caption{Signal current absorption profile 
for $\lambda_0=0$ (no interaction between CPB and NR),
$\Delta_c=0$ (driven exactly on resonance) and $\Omega_c=3$ (by a strong control field). The negative values of the absorption profile represent stimulated emission, i.e., amplification of the signal current. Other parameters used are
$\gamma_0=5\times10^{-4}$, and $(T_2/T_1)=0.2$.}
\label{fig:Mollow_amplification}
\end{figure}

Note that the decay rates of levels $|+,N_+\rangle$ are much larger than those of levels $|-,N_-\rangle$ for arbitrary values of $N$ although they have roughly the same decoherence rates. This is because $|-\rangle_z=|g\rangle$ is assumed to be the lowest electronic energy state of the CPB qubit and the phonon number decay rate $\gamma$ is taken to be much smaller than $T_1^{-1}$ and $T_2^{-1}$ ($\gamma T_2=\gamma_0=5 \times 10^{-4}$, and $T_2/T_1=0.2$). Due to these decay and decoherence rates, the condition \cite{Fleischhauer05} for which all the important features of the RAIT remains considerably observable requires $\Omega^2\gg (T_1 T_2)^{-1}$ or equivalently $\Omega_c^2\gg (T_2/T_1)$.   For the parameters chosen in Figs.~\ref{fig:absorption_Omega} and \ref{fig:absorption_lambda} for our CPB-NR system, the condition is well satisfied. If, however, the value of $\omega_c$ decreases and the value of the ratio $(T_2/T_1)$ increases, then the EIT absorption hole (dip) will become shallow.
In a two-level atomic system driven by a strong, resonant field, when the Rabi frequency of the driving field is greater than the atomic decay rate, the resonance fluorescence spectrum exhibits three-peak structure, called Mollow three-peak spectrum \cite{Mollow69,Scully97}. If we set $\lambda_0=0$ (no interaction between the CPB qubit and the NR) and $\Delta_c=0$ (i.e., the control current resonantly interacting with the CPB qubit) with $\Omega_c$ much larger than the decay rate of the CPB qubit, a Mollow three-peak spectrum but with different relative peak widths compared to a two-level atomic system could be expected in our CPB qubit system. The difference in the peak widths 
lies on the fact that usually $T_2/T_1=2$ in an atomic system [see Eqs.~(\ref{T2}) and ({\ref{T1}) if the pure dephasing rate $\gamma_2=0$], while a typical value of $T_2/T_1=0.2$ is chosen for the CPB qubit system. If we apply a weak probe field (signal current) and calculate the signal-current absorption power profile, we find that the signal-current absorption power profile takes on the negative values (see Fig.~\ref{fig:Mollow_amplification}), representing stimulated emission rather than absorption \cite{Mollow72,Mollow77}. This amplification of the signal current may be understood to happen primarily at the expense of the strong driving (control) current, which experiences an increased attenuation rate. Similar amplification of the probe-field profile in a strongly driven two-level atomic system at optical frequencies was predicted in Ref.~\cite{Mollow72} and experimentally observed and verified in Ref.~\cite{Mollow77}.

The essential point of RAIT in our system is that the absorption power of
the signal current goes abruptly to zero (almost) for
$\omega_s=\omega_c+\omega_r$. This gives us a method to measure
the vibration frequency of the NR with high precision. The
procedure is as follows. Fixing the frequency of the control
current which is close to the resonant frequency of the CPB, and
changing the frequency of the signal current, when the signal
current becomes transparent, obviously we have
$\omega_s=\omega_c+\omega_r$, i.e., the difference between the
frequency of the control current and the signal current is the
frequency of the NR, $\omega_r$.
One can also see from Fig.~\ref{fig:absorption_lambda} that the width of the
absorption hole (valley)
increases with the increasing values of $\lambda_0$. This can be understood as follows. With the increase of the coupling strength $\lambda_0$, the
absorption peak shifts to the right, and the minima of the non-absorption hole (valley) appears however at $\Delta_s=0$ (as we have chosen $\Delta_c=\omega_0$ mentioned above). As a result, the width of the absorption hole (valley) increases.

In Fig.~\ref{fig:absorption_decay}, we draw the absorption power as a function of the NR decay rate $\gamma_0$ for $\Delta_s=0$.
Again, the absorption power is scaled in unit of
$P_0$ which is the maximum value of $P$ for the parameters chosen
in Fig.~\ref{fig:absorption_decay}. It shows that the absorption
power of the signal current increases with the increase of the
decay rate of the NR. In this way, we can determine the decay rate
of the NR and investigate its relation with the environmental
temperatures according to the absorption power of the signal
current.

\begin{figure}[htbp]
\begin{center}
\includegraphics  [height=5cm,width=9cm] {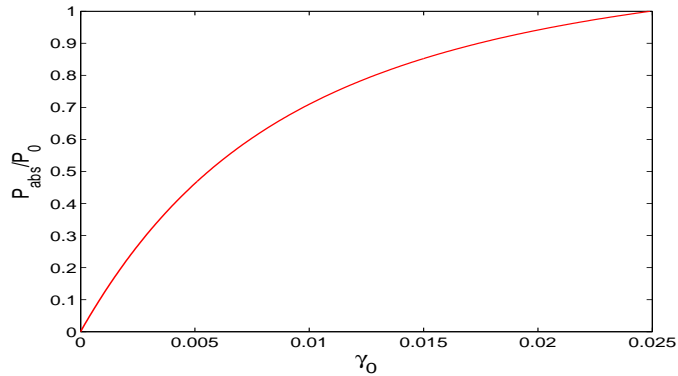}
\end{center}
\caption{Scaled absorption power of the signal current as a function of the NR decay rate $\gamma_0$ for $\Delta_s=0$. Other parameters used are
$\Omega_c=3$, $\lambda_0=2.5\times10^{-3}$, and $\Delta_c=\omega_0=50$.}
\label{fig:absorption_decay}
\end{figure}

\section{Conclusion}
 In conclusion, we have demonstrated the
EIT phenomena in a system of a CPB qubit coupled to a NR. Though the CPB is an
effective two-level system, the NR contributes additional
auxiliary energy levels so that the EIT phenomena can be realized in such a
system. Without the NR, the EIT phenomena will disappear. So we
call our scheme resonator-assisted induced transparency (RAIT). Our proposal
which is within the reach of current experimental technology
provides a detection scheme to measure physical properties, such
as the vibration frequency and the decay rate of the coupled NR,
or the decay and decoherence rates of the CPB qubit, if one set
of the values of either the CPB or NR properties is known by other means. 
We have 
specifically demonstrated here a case to measure the vibration
frequency and the decay rate of the NR using the RAIT technique.
Such a high contrast and high accuracy NR frequency measurement could
potentially provide an alternative way to other NR measurement
schemes, such as magnetomotive detection.

\section*{Acknowledgments}
This work has been supported in part by the National Natural Science
Foundation of China and the National Minister of Education Program
for Changjiang Scholars and Innovative Research Team in University
(PCSIRT). HSG would like to acknowledge support from
the National Science Council, Taiwan under grant numbers NSC96-2112-M-002-007 
and NSC97-2112-M-002-012-MY3, 
support from the National Taiwan University, Taiwan under grant number 
97R0066-67, and support from the focus group program of the National Center for
Theoretical Sciences, Taiwan. XZY acknowledges support
from the National Natural Science Foundation of China under grant
number 10874117. KDZ acknowledges support from the National Natural Science Foundation of China under grant number 10774101.

\end{document}